# $Cs_2InAgCl_6$: A new lead-free halide double perovskite with direct band gap

George Volonakis,[†] Amir Abbas Haghighirad,[‡] Rebecca L. Milot,[‡] Weng H. Sio,[†] Marina R. Filip,[†] Bernard Wenger,[‡] Michael B. Johnston,[‡] Laura M. Herz,[‡] Henry J. Snaith,[‡] and Feliciano Giustino[*,†]

[†]Department of Materials, University of Oxford, Parks Road OX1 3PH, Oxford, UK
[‡]Department of Physics, Clarendon Laboratory, University of Oxford, Parks Road, Oxford OX1 3PU, United Kingdom

E-mail: feliciano.giustino@materials.ox.ac.uk

Phone: (+44) 01865 272380

**Abstract**

$A_2BB'X_6$ halide double perovskites based on bismuth and silver have recently been proposed as potential environmentally-friendly alternatives to lead-based hybrid halide perovskites. In particular, $Cs_2BiAgX_6$ (X = Cl, Br) have been synthesized and found to exhibit band gaps in the visible range. However, the band gaps of these compounds are indirect, which is not ideal for applications in thin film photovoltaics. Here, we propose a new class of halide double perovskites, where the $B^{3+}$ and $B^+$ cations are $In^{3+}$ and $Ag^+$, respectively. Our first-principles calculations indicate that the hypothetical compounds $Cs_2InAgX_6$ (X = Cl, Br, I) should exhibit direct band gaps between the visible (I) and the ultraviolet (Cl). Based on these predictions, we attempt to synthesize $Cs_2InAgCl_6$ and $Cs_2InAgBr_6$, and we succeed to form the hitherto unknown double perovskite $Cs_2InAgCl_6$. X-ray diffraction yields a double perovskite structure with space group F m3m. The measured band gap is 3.3 eV, and the compound is found to be photosensitive and turns reversibly from white to orange under ultraviolet illumination. We also perform an empirical analysis of the stability of $Cs_2InAgX_6$ and their mixed halides based on Goldschmidt's rules, and we find that it should also be possible to form $Cs_2InAg(Cl_{1-x}Br_x)_6$ for x < 1. The synthesis of mixed halides will open the way to the development of lead-free double perovskites with direct and tunable band gaps.

During the last four years Pb-based halide perovskites have revolutionized the field of solution processable solar cells, achieving record power conversion efficiencies above 22%, and surpassing polycrystalline and thin-film silicon photovoltaics (PV).[1–4] While the efficiency of these materials improves steadily, there are two remaining challenges that need to be addressed in order to use perovskite solar cells for electricity production, namely the compound stability and the presence of lead.[5] On the front of Pb replacement, several lead-free perovskites and perovskite-derivatives have been proposed during the past two years as potential substitutes for $MAPbI_3$ (MA = $CH_3NH_3$),[6] including vacancy-ordered double perovskites, cation-ordered double perovskites (also known as elpasolites), and two-dimensional perovskites.[7–17] Among these compounds, Pb-free halide double perovskites based on Bi and Ag were recently proposed as stable and environmentally-friendly alternatives to $MAPbI_3$.[12–15] While these Bi/Ag double perovskites exhibit band gaps in the visible (from 1.9 to 2.2 eV, Ref. 15), the gaps are indirect, which is not ideal for thin film PV applications. In fact indirect band gaps imply weak oscillator strengths for optical absorption and for radiative recombination, therefore indirect semiconductors such as silicon need a much thicker absorber layer, which can be problematic if charge carriers mobilities are low. In order to circumvent this limitation, in this work we develop a new class of Pb-free halide double perovskites, which exhibit direct band gaps. We first use available databases and first-principles calculations to identify double perovskites with direct band gaps, and then we synthesize and characterize the new In-based compound $Cs_2InAgCl_6$. X-ray diffraction (XRD) indicates a double perovskite structure with space group $Fm\bar{3}m$ at room temperature, with a direct band gap of 3.3 eV.

The starting point of our investigation is the recent work on $Cs_2BiAgCl_6$ by some of us, Ref. 12. In that work we succeeded in synthesizing the new double perovskite $Cs_2BiAgCl_6$ by following the well-known reaction route for making $Cs_2BiNaCl_6$,[18] and by replacing the NaCl precursor by AgCl. The successful replacement of Na by Ag can be attributed to the close match between the ionic radii of $Na^+$ (1.02 Å) and of $Ag^+$ (1.15 Å). Motivated by this observation, here we set to discover new halide double perovskites containing Ag. To this aim we search for known halide elpasolites containing $Na^+$ as the $B^+$ cation. We consider data from the Materials Project (www.materialsproject.org), the International Crystal Structure Database (ICSD), and literature reviews.[18–21] From Refs. 18–21 we find 42 elpasolites with composition $Cs_2B^{3+}NaX_6$, X = Cl, Br, I. In 30 of these compounds the $B^{3+}$ cation is a lanthanide or actinide; in 7 compounds we find transition metals with small ionic radii (<0.9 Å), namely Sc, Ti, Y, and Fe; and 5 compounds contain Bi, Sb, Tl, or In. Lanthanides, actinides and transition metals pose a challenge to computational predictions based on DFT, therefore we leave them aside for future work. Bi and Sb double perovskites were extensively discussed in Ref. 12. The incorporation of Tl in halide double perovskites was proposed in Ref. 22, however this element is highly toxic and hence unsuitable for optoelectronic applications. Indium, on the other hand, is a common element in the optoelectronic industry, for example it is used to make transparent conductors in the form of tin doped indium oxide (ITO). Two In/Na-based halide double perovskites have been reported so far, $Cs_2InNaCl_6$ [19,20] and $Cs_2InNaBr_6$.[20] Following this observation, we proceed to assess the structural, electronic, and optical properties of the hypothetical compounds $Cs_2InAgX_6$ with X = Cl, Br, I using DFT calculations.

As a structural template for our calculations we use the face-centered cubic Fm-3m elpasolite unit cell, which contains one $B^+X_6$ and one $B^{3+}X_6$ octahedra.[12,14] The resulting atomistic model consists of $InX_6$ and $AgX_6$ octahedra which alternate along the [100], [010], and [001] directions, as shown in Figure 1a. This rock-salt ordering of the B-site cations is also found in many oxide double perovskites. After optimizing the structure within the local density approximation (LDA) to DFT (see Methods), we obtain the lattice constants a = 10.20 Å, 10.74 Å, and 11.52 Å for X = Cl, Br, and I, respectively. The calculated lattice constants are slightly smaller than what we found for the corresponding Bi-based double perovskites (10.50-11.76 Å).[12] This can be attributed to the smaller size of $In^{3+}$ as compared to $Bi^{3+}$. Despite the presence of two different B-site cations, within DFT/LDA the $BX_6$ octahedra are of similar size; for example in the case of $Cs_2InAgCl_6$ the In-Cl and Ag-Cl bond lengths are 2.50 Å and 2.59 Å, respectively. Employing higher level non-local functionals for the structural optimization yields a slightly larger lattice constant, 10.62 Å for $Cs_2InAgCl_6$, and bond lengths of 2.54 Å and 2.77 Å for In-Cl and Ag-Cl, respectively. All parameters related to structural optimization with different functionals are included in Table S1 of the Supporting Information.

A complete study of the stability of these hypothetical compounds would require the calculation of the phonon dispersion relations (dynamical stability) and of the quaternary convex hulls (thermodynamic stability).[23] Since these calculations are demanding, we perform a preliminary assessment of compound stability using Goldschmidt's empirical criteria. According to these criteria, one can assign a perovskite with formula $ABX_3$ two parameters called the tolerance factor and the octahedral factor. The tolerance factor is defined as $t = (R_A + R_X)/\sqrt{2}(R_B + R_X)$, with $R_A$, $R_B$, and $R_X$ the ionic radii of the elements in a $ABX_3$ perovskite; stable structures usually correspond to $0.75 < t < 1.0$.[24] The octahedral factor is defined as $\mu = R_B/R_X$, and stable structures tend to have $\mu > 0.41$. In order to evaluate these parameters we employ the Shannon ionic radii. Since we have double perovskites, and therefore different radii for the $B^+$ and $B^{3+}$ sites, as a first attempt we consider for $R_B$ the average between the radii of $In^{3+}$ and $Ag^+$. For $Cs_2InAgX_6$ with X = $Cl^-$, $Br^-$, $I^-$ we obtain ($\mu$, t) = (0.54, 0.94), (0.50, 0.93), and (0.44, 0.91), respectively. All these parameters fall within the range of stability of standard halide perovskites,[24] therefore they are not very informative. A more stringent test can be obtained by considering the limiting cases where the structures were composed entirely of $InX_6$ octahedra or of $AgX_6$ octahedra. In this scenario we find that the tolerance factor is within the stability region for all compounds ($0.86 < t < 1.0$), but the octahedral factor $\mu$ is outside of this region for X = Br and I. The evolution of the octahedral factor as one moves from $Cs_2InAgCl_6$ to $Cs_2InAgI_6$ is shown in Figure 1b. It is apparent that the onset of the instability relates to the $InX_6$ octahedra, and in particular to the fact that the ionic radius of $In^{3+}$ is too small ($R_{In^{3+}} = 0.8 °A$) to coordinate 6 $I^-$ anions ($R_{I^-} = 2.2 °A$). This analysis, albeit very crude, suggests that the only possible In/Ag halide double perovskites that may be amenable to synthesis are those with Cl, Cl/Br mixes, and possibly Br.

Guided by these results, we investigate the electronic structure of $Cs_2InAgCl_6$. In order to overcome the well-known limitations of DFT/LDA for the description of band gaps, we employ the HSE and the PBE0 hybrid functionals (see Methods). We also checked that, unlike in $MAPbI_3$,[25] spin-orbit effects are not important for In/Ag double perovskites, see Figure S1 of the Supporting Information. The band structure of $Cs_2InAgCl_6$ is shown in Figure 2a. The most striking finding is that the band gap is direct, with both valence and conduction band extrema at the center of the Brillouin zone ($\Gamma$ point). The band gap is rather sensitive to the functional employed for the structural optimization (see Table S1), and ranges from 2.1 eV (LDA structure, HSE gap) to 3.3 eV (HSE structure, PBE0 gap). Given this uncertainty, we choose to report $2.7 \pm 0.6$ eV as the nominal band gap from the calculations. In Table 1 we show the previously calculated[12,15] and measured[12–15] band gaps for the case of $Cs_2BiAgX_6$ (X = Cl, Br). The measured band gaps are within the HSE-PBE0 range, hence we expect a similar trend for $Cs_2InAgCl_6$.

In Figure 2b we show the square modulus of the electronic wavefunctions at the band edges. The bottom of the conduction band is mainly comprised of Cl-3p and In-5s/Ag-5s states, while the top of the valence band is derived from Cl-3p and In-4d/Ag-4d states. A detailed projected density of states and a qualitative molecular orbital diagram of $Cs_2InAgCl_6$ are shown in Figure S2 of the Supporting Information. The valence band top of $Cs_2InAgCl_6$ has no occupied In-s orbitals. This is closely linked with the direct character of the band gap of this compound. In fact, in the related Bi/Ag elpasolite $Cs_2BiAgCl_6$,[12,15] the Bi s states at the top of the valence band interact with the directional Ag d states along the [100] direction; this interaction pushes the valence band top to the X point and leads to an indirect band gap in $Cs_2BiAgCl_6$. To confirm this point we artificially removed the Bi s states from the top of the valence band of $Cs_2BiAgCl_6$, using a fictitious Hubbard U correction of 10 eV. The resulting band structure, shown in Supporting Figure S3, exhibits a direct gap at $\Gamma$, as in the case of the present compound $Cs_2InAgCl_6$.

The electron effective mass calculated for $Cs_2InAgCl_6$ is $m^*_e = 0.29m_e$ ($m_e$ is the free electron mass); the hole effective mass is $m^*_h = 0.28\ m_e$. In the calculation of the hole effective mass we have not considered the non-dispersive band which can be seen in Figure 2a between $\Gamma$ and X. This flat band originates from the hybridization of Cl-$3p_{x,y}$ states and $4d_{x^2-y^2}$ states of In and Ag, which leads to a two-dimensional wavefunction confined within the equatorial (001) plane (wavefunction labelled as 'iii' in Figure 2b). This flat band will likely prevent carrier transport along the six equivalent X wavevectors, and possibly favor the formation of deep defects.

Since we expect $Cs_2InAgCl_6$ to be stable, and to exhibit a direct band gap as well as small and balanced effective masses, we proceed to attempt the synthesis of this compound. We prepare samples of $Cs_2BiAgCl_6$ by precipitation from an acidic solution of hydrochloric acid. A mixture of a 1 mmol $InCl_3$ (Sigma Aldrich, 99.99%) and AgCl (Sigma Aldrich, 99%) is first dissolved in 5 mL 10 M HCl. Then 2 mmol of CsCl (Sigma Aldrich, 99.9%) are added and the solution is heated to 115 °C. A white precipitate forms immediately after adding CsCl. We leave the hot solution for 30 min under gentle stirring, to ensure a complete reaction before filtering. We wash the resulting solid with ethanol and dry in a furnace overnight at 100 °C. The as-formed powder appears stable under ambient conditions. In Figure 3a we show the measured X-ray diffraction (XRD) pattern of this powder. The refined lattice parameter is 10.47 Å, only 2.6% larger than in our DFT/LDA calculation. This is in line with the typical underestimation of lattice parameters by DFT/LDA in the range of 1-2%. The refined structural parameters are provided in Table S2 of the Supporting Information. In Figure 3a we include the powder diffraction pattern calculated from our optimized DFT/LDA structure. We can see that all the peaks in the calculation and in the XRD measurement match very closely; there is a small offset between the two patterns, which is related to the difference in the lattice parameters. To confirm this point, we re-optimize the theoretical model using the experimental lattice constant; in this case we find no discernible difference between the two patterns, as shown in Figure S4a of the Supporting Information. This comparison indicates that the synthesized compound is a double perovskite or elpasolite, within a $F\bar{m}3m$ space group. We also measured the XRD pattern after exposing the samples to ambient conditions for more than three months, as shown in Figure S5 of the Supporting Information. The compound appears to be very stable with no structural decomposition observed. While we can successfully synthesize $Cs_2InAgCl_6$, our preliminary attempts at making the related bromide compound, $Cs_2InAgBr_6$, have been unsuccessful thus far. In fact, by following a similar route as above, with CsBr instead of CsCl, we obtain a pale yellow powder that does not match the elpasolite Fm3m crystal structure.

In order to investigate whether the In and Ag cations are fully ordered, we have calculated the

XRD pattern in three different scenarios: (1) using an ordered structure optimized within DFT/LDA, (2) using an ordered structure optimized with DFT/HSE, (3) using the measured lattice parameter and a perovskite structure with 0.5 In and 0.5 Ag occupations on the Wyckoff sites. The scenarios (1) and (2) are meant to describe a fully-ordered $Cs_2InAgCl_6$ crystal using DFT functionals which yield rather different In-Cl and Ag-Cl bond lengths (as shown in Table S1). These calculated XRD patterns are compared to experiment in Figure S6 of the Supporting Information. The main peaks associated with ordering effects are indicated by the green arrows in this figure. We find that the main peaks associated with ordering effects (indicated by the green arrows) are extremely weak even in the theoretical fully-ordered structures. Furthermore, we see that in the HSE structure these peaks are slightly more visible than in the LDA structure. This effect is due to the more pronounced difference in bond lengths between In-Cl and Ag-Cl in the HSE structure (as seen in Table S1). In the fully-disordered model the intensity of the ordering peaks is vanishing. The difference between the XRD patterns calculated for the ordered structures (Figure S6b and S6c) and the disordered structure (Figure S6d) are much smaller than our experimental resolution (Figure S6a). Therefore, in the case of $Cs_2InAgCl_6$, XRD is not sufficient to probe the nature of cation order/disorder. Given this uncertainty on the structure, we explored the effects of disorder on the electronic structure. To this aim we considered a virtual crystal approximation with the In and Ag sites replaced by the pseudo-atom 0.5 In + 0.5 Ag. As shown in Figure S7, the DFT/HSE band structure of this disordered model is very close to that of the ordered model, with the same direct band gap at $\Gamma$. This demonstrates that our electronic structure analysis should be mostly insensitive to cation ordering.

In order to characterize the optical properties of $Cs_2InAgCl_6$, we measured the diffuse reflectance and applied the Kubelka-Munk theory (see Methods) to estimate the absorbance, and the time-resolved photoluminescence (PL) spectra. The Kubelka-Munk function F (R) is shown in Figure 3b. We can clearly recognize the onset of absorption near 380 nm (3.3 eV) and further identify a second absorption feature at 585 nm (2.1 eV). The full reflectance spectra is shown in Figure S8 of the Supporting Information. In Figure 3c we see a well-defined PL peak centered around 608 nm (2.04 eV), with a FWHM of 120 nm (0.37 eV).This peak appears to redshift on a timescale of 100 ns, suggesting that subgap states are being filled on this timescale.

A compound with the absorption onset and PL peak close to 2.1 eV should be colored orange and not white as the $Cs_2InAgCl_6$ powder. However, we have observed that under photo-excitation the compound turns orange as shown in the insets of Figure 3b. This coloration might relate to local photo-induced electronic or structural changes. For example, the precursor compound AgCl is a known photosensitive material, where the oxidation state of Ag changes upon illumination. This photochromic behavior may be connected with the strong sensitivity of the electronic structure of $Cs_2InAgCl_6$ to small changes in the Ag-Cl and In-Cl bond lengths, as shown in Table S1 of the Supporting Information. The process observed here is fully reversible and upon return to ambient light the sample becomes immediately white. The orange coloration is consistent with the optical absorption around 2.1 eV and the emission spectra shown in Figure 3b and c. Based on these observations we propose that the measured absorption and emission around 2.1 eV might relate to (possibly photo-induced) defect states, and the actual band gap of the compound is at 3.3 eV. In Figure 3d we see that the time-resolved PL intensity of $Cs_2InAgCl_6$ exhibits two timescales, a fast relaxation with a lifetime of 1 ns, followed by a very slow decay with a long lifetime of 6 μs. This observation is consistent with time-resolved PL measurements on the related $Cs_2BiAgX_6$ (X = Cl, Br) double perovskites, which also exhibit two decay components with very different lifetimes.[13] For completeness in Figure S4b of the Supporting Information we show that the PL lifetime is insensitive to the intensity of the pump laser. Taken together, the PL redshift with time, and the fast initial PL decay component suggest that a tail of subgap defect states or energetic disorder is present. We now employ DFT to compare the ideal optical absorption spectrum of $Cs_2InAgCl_6$ and $Cs_2InAgBr_6$ with the spectra calculated for the standard semiconductors Si and GaAs,[26] and for $MAPbI_3$ (unpublished results). In Figure 4a we see that, while $Cs_2InAgCl_6$ is a direct-gap semiconductor, the absorption coefficient of the perfect crystal is considerably smaller than those of Si, GaAs, and $MAPbI_3$ throughout the visible spectrum. On the other hand, if we could synthesize an ideal crystal of the bromine compound, $Cs_2InAgBr_6$, we should obtain an absorption coefficient which is comparable or even higher than in silicon.

In view of developing more efficient absorbers based on the In/Ag combination, we theoretically investigate the optical properties of mixed halide compounds. To this aim we consider the hypothetical solid solutions $Cs_2InAg(Cl_{1-x}Br_x)_6$, $Cs_2InAg(Br_{1-x}I_x)_6$ and $Cs_2InAg(Cl_{1-x}I_x)_6$ with x = 0.25, 0.50, and 0.75. To model mixed-halide double perovskites without inducing artificial ordering effects, we employ the virtual-crystal approximation (see Methods). Figure 4b shows a ternary map of the band gap calculated for these mixes, as obtained by interpolating linearly from the above combinations. We report band gaps calculated using DFT/HSE and DFT/PBE0 for $Cs_2InAgCl_6$. The band gaps of the mixed halide double perovskites are tunable within the visible spectrum. These results indicate that the the In/Ag halide double perovskites may be promising for applications in tandem solar cell architectures. [4]

In summary, by combining first-principles calculations and experiments we discovered a novel direct band gap halide double perovskite, $Cs_2InAgCl_6$. This new compound crystallizes in a double perovskite structure with space group Fm-3m, has a direct band gap at Γ of 3.3 eV. In addition, this new compound exhibits PL at around 2.1 eV and shows an interesting photochromic behavior whereby the color changes reversibly to orange under UV illumination. Our analysis indicates that by developing mixed halides $Cs_2InAgX_6$ with X = Cl, Br, and I, it should be possible to obtain good optical absorbers with tunable and direct band gaps. The existence of the related compound $Cs_2InNaBr_6$ (with Na in place of Ag) strongly suggests that it should be viable to synthesize also $Cs_2InAgBr_6$. Overall, we expect that the exploration of mixed-halide double perovskites starting from $Cs_2InAgCl_6$ will offer new opportunities for developing environmentally friendly perovskite photovoltaics and optoelectronics.

Computational Methods: Structural optimization was carried out within DFT using the Quantum ESPRESSO suite.[27] Ultrasoft pseudopotentials[28,29] were used to describe the electron-ion interaction, and exchange-correlation effects were taken into account within the LDA.[30,31] The planewaves kinetic energy cutoffs were set to 35 Ry and 280 Ry for the electronic wavefunctions and the charge density, respectively. The Brillouin zone was sampled using a 15×15×15 unshifted grid. The thresholds for the convergence of forces and total energy were set to $10^{-3}$ Ry and $10^{-4}$ Ry, respectively. In order to overcome the DFT band gap problem we employed hybrid functional calculations as implemented in the VASP code.[32] We used the PBE0 functional[33] and the HSE06 functional.[34] In the latter case we set the screening parameter to 0.2 $Å^{-1}$ and we used a mixing of 25% of Fock exchange with 75% of PBE exchange. In the hybrid calculations we used the projector augmented wave method,[35] with a 400 eV kinetic energy cutoff. In the hybrid calculations the Brillouin zone was sampled using a 6×6×6 unshifted grid for Cl or Br. For the I-based compound, $Cs_2InAgI_6$, DFT/LDA yields a spurious band crossing and the resulting band structure is metallic. In this case, in order to perform hybrid calculations we sampled the Brillouin zone using a 2×2×2 shifted grid. We verified the correctness of the results by repeating the same procedure for $Cs_2InAgCl_6$. For the calculations of the optical absorption coefficients we used the YAMBO code[36] within the random-phase approximation. In this case we used norm-conserving pseudopotentials[37] (with kinetic energy cutoff of 120 Ry), the PBE[38] generalized gradient approximation, and a dense Brillouin zone grid with 40×40×40 points. In order to correct for the band gap underestimation in DFT/PBE we used scissor corrections as obtained from our HSE/PBE0 calculations. The optical f-sum rule was maintained by scaling the oscillator strengths, following Ref. 39. The calculations based on the virtual-crystal approximation were performed by using the procedure of Ref. 40.

Experimental Methods: Powder XRD was performed using a Panalytical X'pert diffractometer (Cu-K$\alpha_1$ radiation; λ = 154.05 pm) at room temperature. Structural parameters were obtained by Rietveld refinement using the General Structural Analysis Software.[41,42] A Varian Cary 300 UV-

Vis spectrophotometer with an integrating sphere was used to acquire diffuse reflectance. To estimate the visible light absorption spectrum, we apply the Kubelka-Munk theory to convert the diffuse reflectance in the F (R) function, $F(R) = (1 - R)^2/2R = K/S$, where R is the absolute reflectance of the sample, K is the molar absorption coefficient, and S is the scattering coefficient. Time-resolved PL spectra were recorded following laser excitation at 405 nm, at a repetition rate of 10 MHz (Picoquant, LDH-D-C-405M). The PL signal was collected and directed toward a grating monochrometer (Princeton Instruments, SP-2558), and detected with a photon-counting detector (PDM series from MPD).

## Acknowledgement

The authors are grateful to Marios Zacharias for sharing the calculated absorption coefficients of Si, GaAs and $MAPbI_3$. The research leading to these results has received funding from the Graphene Flagship (Horizon 2020 grant no. 696656-GrapheneCore1), the Leverhulme Trust (Grant RL-2012-001), the UK Engineering and Physical Sciences Research Council (Grant No. EP/J009857/1, EP/M020517/1 and EP/L024667/1). The authors acknowledge the use of the University of Oxford Advanced Research Computing (ARC) facility (http://dx.doi.org/10.5281/zenodo.22558) and the ARCHER UK National Supercomputing Service under the 'AMSEC' Leadership project and PRACE for awarding us access to the Dutch national supercomputer 'Cartesius'.

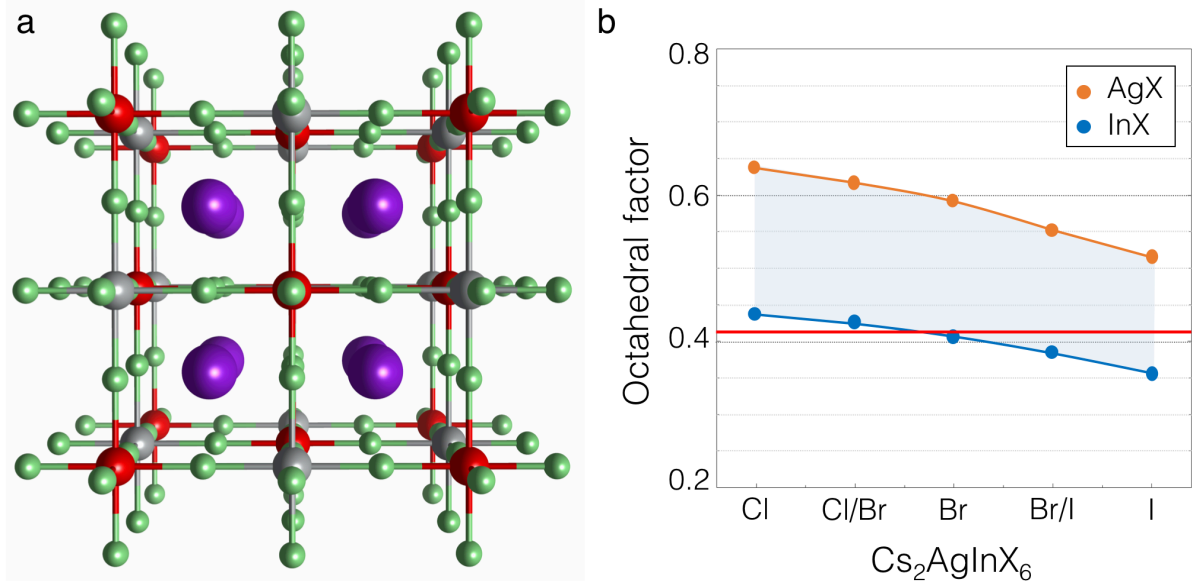

**Figure 1: Atomistic model and octahedral factors of the In-based halide double perovskites $Cs_2InAgX_6$ (X = Cl, Br, I):** (a) Ball-and-stick model of $Cs_2InAgCl_6$, with In atoms in red and Ag atoms in gray. The green spheres indicate Cl, and the purple spheres in the center of each cavity are Cs atoms. The primitive unit cell contains one $InCl_6$ and one $AgCl_6$ octahedra in a face-centered cubic structure, with space group Fm3m; here we show the conventional cubic cell. (b) Octahedral factor $\mu = R_B/R_X$ corresponding to $AgX_6$ octahedra (orange dots), and to $InX_6$ octahedra (blue dots), for solid solutions of Cl, Br, and I. The perovskite structure is expected to be unstable for values of the octahedral factor below $\mu = \sqrt{2}-1 = 0.41$, as indicated by the red horizontal line.[24]

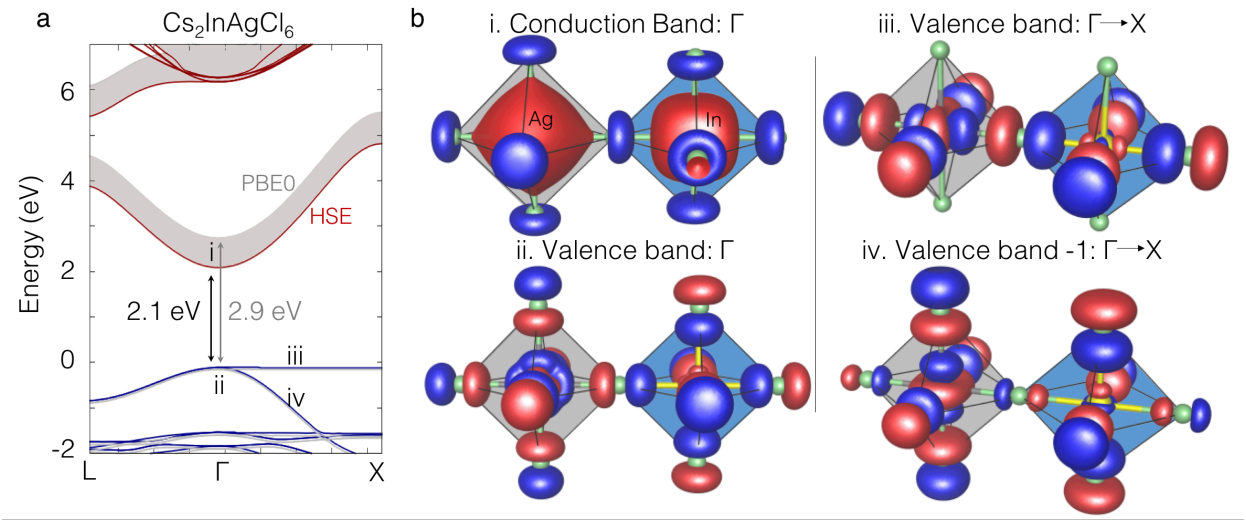

**Figure 2: Electron band structure and square modulus of the wavefunctions for $Cs_2InAgCl_6$:** (a) Band structures and band gaps of $Cs_2InAgCl_6$, as calculated by using the HSE hybrid functional (blue and red lines) or the PBE0 functional (shaded area). The zero of the energy scale is set to the top of the valence band. (b) Isosurface plots of the square modulus of the Kohn-Sham wavefunctions corresponding to (i) the bottom of the conduction band at $\Gamma$; (ii) the sum of the two degenerate states at the valence band top at $\Gamma$; (iii) the highest occupied state along the $\Gamma X$ line, for a wavevector $k \to \Gamma$ (0.2 $\Gamma X$); the second-highest occupied state along $\Gamma X$, for $k \to \Gamma$ (0.2 $\Gamma X$).

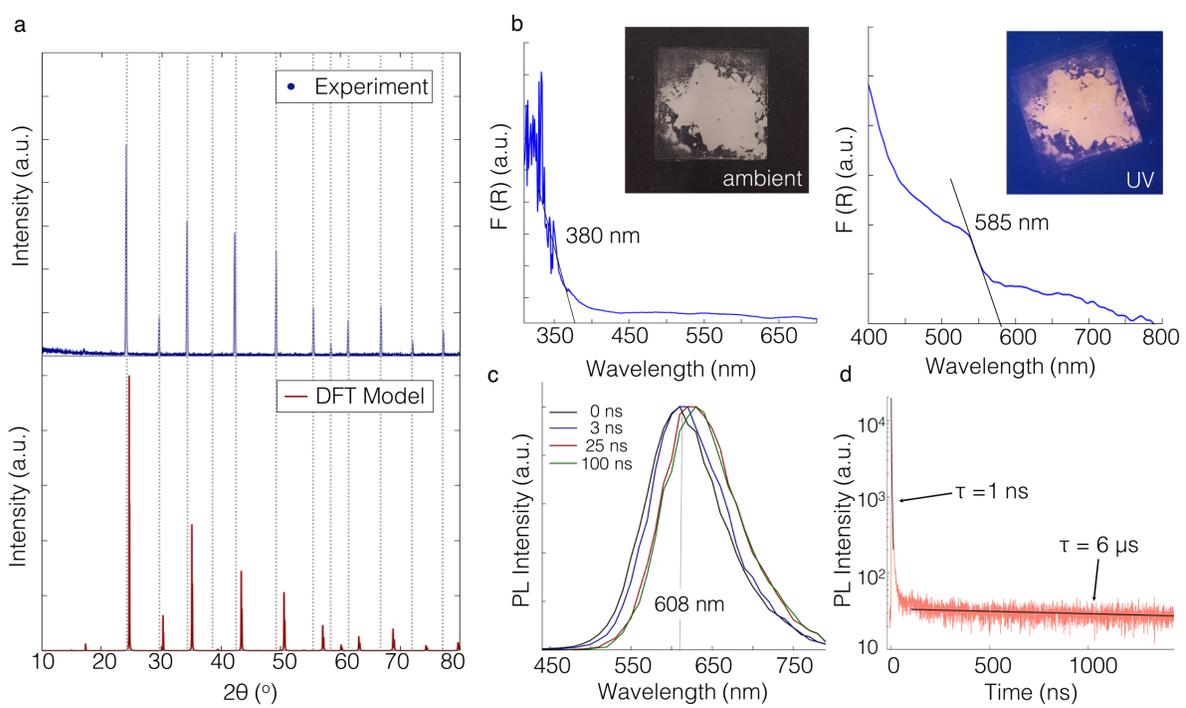

**Figure 3: Structural and optical characterization of $Cs_2InAgCl_6$:** (a) Measured powder XRD pattern for the as-synthesized $Cs_2InAgCl_6$ (top), and the XRD pattern calculated from the atomistic model optimized using DFT/LDA (bottom). (b) Measured UV-Vis absorbance showing the band gap of $Cs_2InAgCl_6$ at 380 nm and the defect-related features at 585 nm as discussed in-text. The straight lines are a guide to the eye. The insets show $Cs_2InAgCl_6$ under ambient light and under UV illumination. (c) Normalized PL spectrum recorded as a function of time following excitation at 405 nm. The vertical line indicates the centre wavelength of the PL at zero time, defined by the arrival time of the excitation pulse on the sample. (d) Time-resolved PL intensity recorded for a powder sample of $Cs_2InAgCl_6$. The fast and slow components of the PL decay are indicated on the plot. The fast (slow) component was fit with a stretched exponential (monoexponential) function, and τ represents the average lifetime.

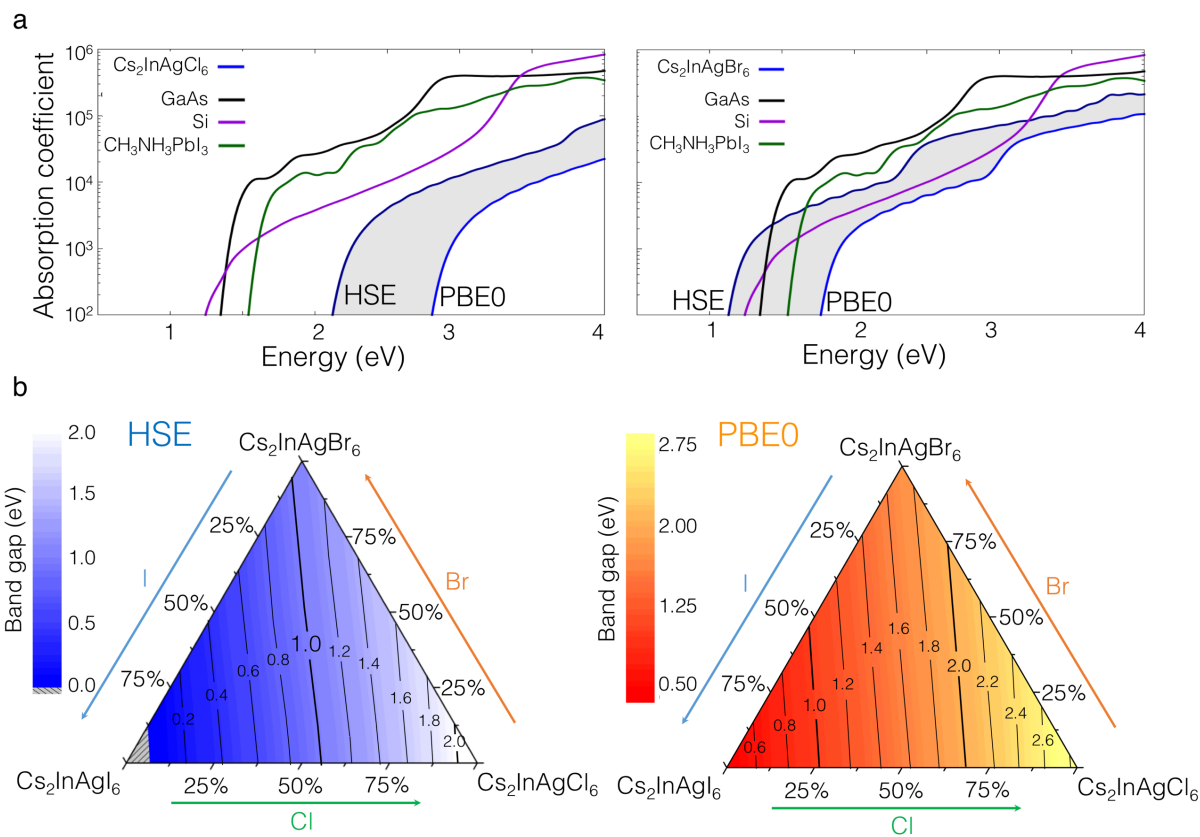

**Figure 4: Theoretical optical absorption coefficient and band gap of mixed halides:** (a) Calculated absorption coefficient of the compound synthesized in this work, $Cs_2InAgCl_6$ (left, blue lines), and of the hypothetical compound $Cs_2InAgBr_6$ (right, blue lines). For comparison we show the theoretical absorption coefficients of silicon (purple), gallium arsenide (black), as calculated in Ref. 26, and $MAPbI_3$ (green) (unpublished results). (b) Calculated band gaps of hypothetical mixed-halide double perovskites $Cs_2InAg(Cl_{1-x-y}Br_xI_y)_6$ within the HSE (left) and PBE0 (right) hybrid functional. The corners of the triangle correspond to $Cs_2InAgX_6$ with X = Cl, Br, I.

**Table 1: Comparison between calculated and measured band gaps of halide double perovskites:** The first column reports our present results for $Cs_2InAgCl_6$ (HSE and PBE0 calculations; the range corresponds to different structural relaxations as detailed in Table S1), the second and third columns report PBE0 calculations and measurements on $Cs_2BiAgCl_6$ and $Cs_2BiAgBr_6$ from Refs. 12, 13, 14 and 15. The HSE calculations are from this work.

|      | $Cs_2InAgCl_6$ | $Cs_2BiAgCl_6$ | $Cs_2BiAgBr_6$ |
|------|----------------|----------------|----------------|
| HSE  | 2.1-2.6        | 2.1            | 1.7            |
| PBE0 | 2.9-3.3        | 2.7            | 2.3            |
| Exp. | 3.3            | 2.2-2.8        | 1.9-2.2        |